\documentclass[%
 reprint,
 amsmath,amssymb,
 aps,
 prb,
floatfix,
]{revtex4-1}

\usepackage{graphicx}% Include figure files
\usepackage{dcolumn}% Align table columns on decimal point
\usepackage{bm}% bold math
\usepackage{multirow}
\usepackage{threeparttable}
\begin{document}

\preprint{APS/123-QED}

%%%%%%%%%%%%%%%%%%%%%%%%%%%%%%%%%%%%%%%%%%%%%%%%%%%%%%%%%%%%%%%%%%%%%%%%%%%%%%%%%%%%%%%%%%%%%%%%%%%%%%%%%%%%%%%%%%%%%%%%%%

\title{Benzene Adsorbed on Metals: Concerted Effect of Covalency and van der Waals Bonding}
\author{Wei Liu$^1$}
\author{Javier Carrasco$^2$}
\author{Biswajit Santra$^{1,3}$}
\author{Angelos Michaelides$^4$}
\author{Matthias Scheffler$^1$}
\author{Alexandre Tkatchenko$^1$}
\email{tkatchenko@fhi-berlin.mpg.de}
\affiliation{$^1$Fritz-Haber-Institut der Max-Planck-Gesellschaft, Faradayweg 4-6, D-14195, Berlin, Germany\\
$^2$Instituto de Cat{\'a}lisis y Petroleoqu{\'i}mica, CSIC, Marie Curie 2, E-28049, Madrid, Spain \\
$^3$Department of Chemistry, Princeton University, Princeton, New Jersey 08544, USA\\
$^4$Thomas Young Centre, London Centre for Nanotechnology and Department of Chemistry, University College London, London WC1E 6BT, United Kingdom}
%\date{\today}

%%%%%%%%%%%%%%%%%%%%%%%%%%%%%%%%%%%%%%%%%%%%%%%%%%%%%%%%%%%%%%%%%%%%%%%%%%%%%%%%%%%%%%%%%%%%%%%%%%%%%%%%%%%%%%%%%%%%%%%%%%

\begin{abstract}
The adsorption of aromatic molecules on metal surfaces plays a key role in condensed matter physics and functional materials.
Depending on the strength of the interaction between the molecule and the surface, the binding is typically classified as either physisorption or chemisorption.
Van der Waals (vdW) interactions contribute significantly to the binding in physisorbed systems, but the role of the vdW energy
in chemisorbed systems remains unclear.
Here we study the interaction of benzene with the (111) surface of transition metals,
ranging from weak adsorption (Ag and Au) to strong adsorption (Pt, Pd, Ir, and Rh).
When vdW interactions are accurately accounted for, the barrier to adsorption predicted
by standard density functional theory (DFT) calculations essentially vanishes, producing a metastable
precursor state on Pt and Ir surfaces.
Notably, vdW forces contribute more to the binding of covalently bonded benzene
than they do when benzene is physisorbed.
Comparison to experimental data demonstrates that some of the recently developed methods for including vdW interactions in DFT
allow quantitative treatment of both weakly and strongly adsorbed aromatic molecules on metal surfaces, extending the already
excellent performance found for gas-phase molecules.
\end{abstract}

\maketitle

\section{INTRODUCTION}
The adsorption of aromatic molecules at transition-metal surfaces is important for fundamental and
applied surface science studies,~\cite{tautz2007,gomez2001,jenkins2009}
and these systems show promise as components in (opto)-electronic devices.~\cite{MRS-Kronik}
In the case of weak overlap of electron orbitals between the adsorbate and the substrate surface, the ubiquitous van der Waals (vdW)
interactions is frequently the only force that binds the molecule to the surface. This situation is typically referred to as physisorption.
In the chemisorption case, the covalent or ionic bonding dominates and the effect of vdW interactions on the overall strength of adsorption
is typically assumed to be weak.
In this study, we challenge this conventional view, by demonstrating the significantly larger contribution of vdW energy
to the stabilization of strongly adsorbed benzene on (111) surfaces of Pt, Pd, Rh, and Ir metals when compared to physisorption on Ag(111) and Au(111) surfaces.

Whereas the role of vdW forces in the binding of atoms and molecules in the gas phase is reasonably well understood,
at solid surfaces our understanding remains far from complete. Indeed, until recent developments
(see, \emph{e.g.}, Refs.~\cite{dion2004,tkatchenko2009,grimme2010,klimes2010}) for efficiently incorporating the long-range vdW
energy within density functional theory (DFT) calculations it was not possible to determine the role of the vdW energy for extended systems and adsorption processes.~\cite{tkatchenko2010,victor2012,javi2011}
A large majority of previous theoretical work on vdW interactions
mainly focused on weakly bound systems.~\cite{busse2011,hamada2011,kelkkanen2011,vanin2010,toyoda2010,wellendorff2010,abad2011}
Typical examples include benzene (Bz) adsorbed on the Ag(111) and Au(111) surfaces,~\cite{vanin2010,toyoda2010,wellendorff2010,abad2011}
and noble gases on the Cu(111), Ag(111), Pt(111), and Pd(111) surfaces.~\cite{silva2003,chen2011,victor2012,ambrosetti2012}
A unifying aspect of these studies is the observation that the inclusion of vdW interactions into standard DFT
within the generalized gradient approximation (GGA) often brings a large increase in binding, and results in a
much better agreement with experimental adsorption distances and energies. However, the vdW forces can also have a \textit{qualitative} impact on
the adsorption process.
One particularly interesting example was reported by Bl{\"u}gel's group, showing that the vdW forces are the key ingredient to trigger the
binding of pyridine on Cu(110) from physisorption to weak chemisorption.~\cite{atodiresei2009} Mittendorfer~\textit{et al.}~\cite{mittendorfer2011}
reported a novel mechanism for graphene adsorption on Ni(111), where weak covalent and vdW interactions lead to two different minima in the binding curve.
Similar results were found by Li \textit{et al.}.~\cite{li2012}
Another example was shown in our recent work on the isophorone molecule (C$_{9}$H$_{14}$O) at the Pd(111) surface,
which illustrated that the binding structure and the dehydrogenation pathway in this system can be predicted only after
accounting for vdW interactions.~\cite{liu2012}
The vdW interactions were also shown to play a role in the chemisorption of benzene on the Si(100) surface.~\cite{Johnston2008,Hyun-Jung}
However, in this case the vdW-DF method leads to a \emph{smaller} adsorption energy than the pure PBE functional,~\cite{Johnston2008}
while the PBE+vdW method predicts a vdW contribution of 0.5 eV.~\cite{Hyun-Jung}

In this work, we demonstrate the significant concerted effect of covalent bonding and vdW interactions for benzene
interacting with metal surfaces, leading to qualitative changes in the adsorption behavior when vdW
interactions are accurately treated (see Figure 2).
In particular, our calculations predict a metastable precursor state for benzene on Pt(111) in agreement with the
experimental findings,~\cite{ihm2004} along with a peculiar ``phase transition'' behavior of the projected HOMO/LUMO occupations
of the benzene molecule.
Comparison to experimental data demonstrates that recently developed methods for including vdW interactions in DFT~\cite{klimes2010,victor2012}
allow quantitative treatment of both weakly and strongly adsorbed aromatic molecules on metal surfaces, extending the already
excellent performance found for gas-phase molecules.~\cite{klimes2010,tkatchenko2012}

\section{COMPUTATIONAL METHOD}

We used two different vdW-inclusive approaches in the present work: a newly developed PBE+vdW$^{\rm surf}$ method,~\cite{victor2012} as implemented in the FHI-aims all-electron code;~\cite{blum2009} and the optB88-vdW method,~\cite{klimes2010}
as implemented in the VASP code.~\cite{kresse1996,klimes2011} The PBE+vdW$^{\rm surf}$ approach includes screened vdW interactions
(beyond the pairwise atom-atom approximation) to study adsorbates on surfaces,
by a synergetic linkage of the PBE+vdW method~\cite{tkatchenko2009} for intermolecular vdW interactions with
the Lifshitz-Zaremba-Kohn theory~\cite{zaremba1976} for the dielectric
screening within the metal substrate. While the PBE+vdW$^{\rm surf}$ approach leads to accurate results in the asymptotic limit
by construction, it uses a short-range damping function with one adjusted parameter.
The optB88-vdW method is a modified version of the vdW-DF functional,~\cite{dion2004}
by using an empirically optimized optB88-like exchange functional.
Both PBE+vdW$^{\rm surf}$ and optB88-vdW methods can accurately describe intermolecular interactions with mean absolute
relative errors on the order of 9\%~\cite{klimes2010,tkatchenko2012} compared to
coupled-cluster dimer binding energies for the S22 molecular database.
Less is known about the performance of these methods for solids and weakly adsorbed molecules
on surfaces, although encouraging results have been reported for a few condensed matter
systems.~\cite{klimes2011,Zhang-PRL,javi2011,victor2012}
It is important to benchmark the newly developed methods on a wider set of condensed
matter systems, especially because the PBE+vdW$^{\rm surf}$ and optB88-vdW methods are based
on very different approximations.
For comparison purposes, calculations using the vdW-DF functional and its second version (vdW-DF2)~\cite{lee2010} were
also carried out for the Bz/Pt(111) and Bz/Au(111) systems.

The FHI-aims code was employed for the PBE+vdW$\rm ^{surf}$, PBE+vdW, PBE,~\cite{perdew1996} and local-density approximation (LDA) calculations. We used the ``tight'' settings, including the standard numerical atom-centered orbitals (NAO) basis set ``tier2'' for H and C, and ``tier1'' for transition metals. For all structural relaxations, we used a convergence criterion of 10$^{-2}$ eV\AA$^{-1}$ for the maximum final force. Also a convergence criteria of 10$^{-5}$ electrons for the electron density and 10$^{-4}$ eV for the total energy of the system were utilized for all computations.  The scaled zeroth-order regular approximation (ZORA) was applied for treating relativistic effects. Using these settings, the accuracy in determining the binding energy and equilibrium distance is better than 0.01 eV and 0.005 {\AA}, respectively.

For bulk lattice constant calculations, we used a Monkhorst-Pack~\cite{monkhorst1976} grid of 16 $\times$ 16 $\times$ 16 \textit{k}-points. The lattice constants of the bulk metals have been obtained by using the Birch-Murnaghan equation of state fit to DFT cohesive energy curves.~\cite{birch1947} Using the respective lattice constants from each method,
we built up a 6-layer slabs with a (3 $\times$ 3) unit cell, with no reconstruction of Pt(111) and Au(111). Each slab was separated by a 20 {\AA} vacuum.
%Thus, there are 54 metal atoms in each system. The metal surfaces are represented by 6-layer slabs with
The vdW interactions between metal atoms were also considered when performing the relaxations.
We constrained the bottom four metal layers while fully relaxed the molecule and the uppermost two metal layers during geometry relaxations.
For slab calculations, we used a \textbf{$ 6 \times 6 \times 1 $} \textit{k}-points mesh.

%We considered eight high-symmetry adsorption sites
%for benzene (Bz) on the (111) surfaces of transition metals.
%For face-centered cubic (FCC) crystal structures (including Pt, Au, and other metals studied in the present work), Bz can adsorb at four different positions (atop, bri (abbreviation for ``bridge''), fcc, and hcp), on the close-packed surfaces.
%At each position, there exist two orientations of 0$^\circ$ and 30$^\circ$, referring to the angles of the C--C bond
%being rotated with respect to the close-packed metal rows~\cite{koschel1999,berland2009}.

For the calculation of the binding curves, we changed the adsorption height \textit{d} of Bz, which is evaluated relative to the position of the
unrelaxed topmost metal layer. For each structure, we fixed the \textit{z} coordinates of the carbon backbone and the
metal atoms in the bottommost four of the employed six layer surface model.

%\subsection{Calculation details using VASP code}

The VASP code was employed for the optB88-vdW, vdW-DF, and vdW-DF2 calculations. Inner electrons replaced by the projector augmented wave (PAW) method, whilst the
monoelectronic valence electrons were expanded in plane-waves with a \textit{E}$_{\rm{cut-off}}$ = 500 eV. For slab calculations, we used a \textbf{$ 4 \times 4 \times 1 $} \textit{k}-points mesh. For metal supercell, we used a (3 $\times$ 3) unit cell with 6 atomic layers (3 bottom layers fixed to the corresponding bulk optimal position for each method). Dipole correction was applied along the direction perpendicular to the metal surface. Geometry optimizations were performed with a residual force threshold of 0.03 eV\AA$^{-1}$.

%\subsection{Binding energy curves from different methods}

%%%%%%%%%%%%%%%%%%%%%%%%%%%%%%%%%%%%%%%%%%%%%%%%%%%%%%%%%%%%%%%%%%%%%%%%%%%%%%%%%%%%%%%%%%%%%%%%%%%%%%%%%%%%%%%%
%All possible adsorption sites
%\begin{figure}[!]
%\centering
%\includegraphics[width= 12 cm]{Binding_Energy_Curves}
%\caption{\label{fig:wide}  Adsorption energy --\textit{E}$\rm _{ad}$ as a function of the adsorption height \textit{d} for Bz on Pt(111) (a) and on Au(111) (b) from the PBE+vdW$\rm ^{surf}$, optB88-vdW, PBE, and optB88 methods. For each relaxation, the carbon backbone height $d$ from the surface is kept fixed. The experimental binding distances and adsorption energies are indicated by yellow intervals.}
%\end{figure}
%%%%%%%%%%%%%%%%%%%%%%%%%%%%%%%%%%%%%%%%%%%%%%%%%%%%%%%%%%%%%%%%%%%%%%%%%%%%%%%%%%%%%%%%%%%%%%%%%%%%%%%%%%%%%%%%%

%We also compare the PBE+vdW$\rm ^{surf}$ potential energy curves with those obtained by the optB88-vdW functional. As shown in Figure 1,
%the same qualitative trends are found when using the optB88-vdW method, and both vdW-inclusive methods yield good agreement with the experiments.
%It is also discernable that the non-local energies are more attractive in optB88-vdW, because the non-local correlation energy also contains contributions from very short distances, which cannot be clearly associated to the vdW energy.

\section{RESULTS}

The typical strongly bound Bz/Pt(111) system (adsorption energy 1.57-1.91 eV~\cite{ihm2004}) and the typical weakly bound Bz/Au(111) system
(adsorption energy 0.73-0.87 eV~\cite{syomin2001}) are used first to demonstrate our point.
Accurate experimental data is available for both of these systems, enabling direct quantitative verification of
our theoretical calculations. To demonstrate the differences in the adsorption mechanism, we explore the potential-energy surface (PES) for Bz on the Pt(111) and Au(111) surfaces. We place a single Bz molecule
at the eight high-symmetry adsorption sites of the (111) metal surface,~\cite{saeys2002} followed by geometry relaxation.
The adsorption geometries and energies for Bz on Au(111) and Pt(111) at the preferable adsorption site are shown in Fig. 1 and Table I.
Already here one can clearly distinguish
the different nature of bonding for the adsorption of Bz on Pt(111) and Au(111).
Irrespective of the functional used (PBE, PBE+vdW, and PBE+vdW$^{\rm surf}$), the bri30$^\circ$ is the most preferable site for Bz/Pt(111), with an angle of 30$^\circ$
between the C--C and Pt--Pt bonds, see Fig. 1. This result is consistent with previous periodic slab GGA calculations,~\cite{saeys2002,saeys2003,mittendorfer2003,morin2004,gao2008}
as well as low-energy electron diffraction (LEED)~\cite{wander1991} and scanning tunneling microscopy (STM)~\cite{weiss1993} experiments. Moreover, the PES shows a corrugation of 1.33 eV for Bz/Pt(111)
when using PBE+vdW$^{\rm surf}$. In contrast, the PES for Bz/Au(111) is found to be flat, with only 0.04 eV corrugation.
This result further justifies the STM observations that even at a temperature of 4 K, Bz molecules are capable of diffusing over the Au(111) terraces.~\cite{mantooth2007}

%%%%Figure 1 Structure
%%%%%%%%%%%%%%%%%%%%%%%%%%%%%%%%%%%%%%%%%%%%%%%%%%%%%%%%%%%%%%%%%%%%%%%%%%%%%%%%%%%%%%%%%%%%%%%%%%%%%%%%%%%%%%%%%
%Figure1: Two relaxed structures both at the brige-30 site
\begin{figure}[h!]
\includegraphics[width= 8.2 cm]{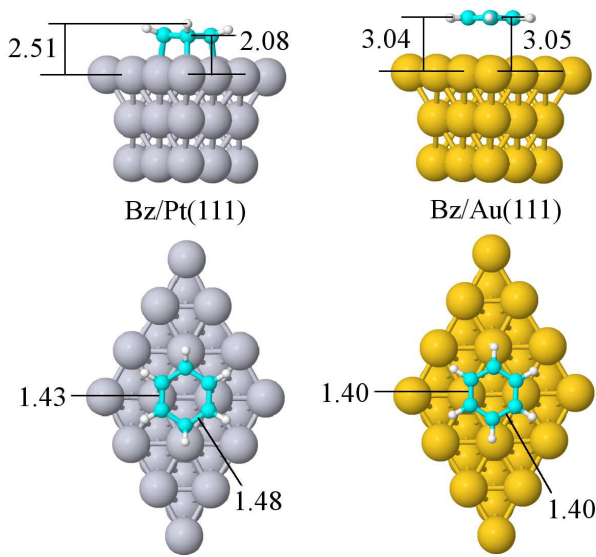}
\caption{\label{fig1:epsart} Adsorption structures of the Bz/Pt(111) system and Bz/Au(111) system, both at the so-called bri30$^\circ$ adsorption site (see text).
 We carried out extended periodic calculations, but only a small part of the supercell is shown. Six metal layers were used but only the topmost three layers are depicted in the figure.
The indicated distances ({\AA}) are obtained based on the PBE+vdW$^{\rm surf}$ optimized structures.
Gray, yellow, cyan, and white spheres represent Pt, Au, C, and H atoms, respectively.
Optimized lattice constants were used for every method.
%The two uppermost metal layers were fully relaxed while the bottom four metal layers were constrained to their bulk positions during relaxation.
}
\end{figure}
%%%%%%%%%%%%%%%%%%%%%%%%%%%%%%%%%%%%%%%%%%%%%%%%%%%%%%%%%%%%%%%%%%%%%%%%%%%%%%%%%%%%%%%%%%%%%%%%%%%%%%%%%%%%%%%%%

%%%%%%%%%%%%%%%%%%%%%%%%%%%%%%%%%%%%%%%%%%%%%%%%%%%%%%%%%%%%%%%%%%%%%%%%%%%%%%Table1: Data for Bz on Pt(111) and Au(111)
%\begin{table}[h!b!p!][wide]
\begin{table}
\centering
\caption{Comparison of adsorption energy (\textit{E}$\rm_{ad}$) and average perpendicular heights
(\textit{d}$\rm_{CM}$ and \textit{d}$\rm_{HM}$ for carbon-metal and hydrogen-metal, respectively) between DFT calculations and
experimental data for Bz on Pt(111) and Au(111). The distances are referenced to the average positions of the relaxed topmost metal atoms.
The adsorption energy \textit{E}$\rm{_{ad}}$ is defined as \textit{E}$\rm{_{ad}}$ =
--(\textit{E}$\rm_{AdSys}$--\textit{E}$\rm_{Me}$--\textit{E}$\rm_{Bz}$), where the subscripts \textit{AdSys},
\textit{Me}, and \textit{Bz} denote the adsorption system, the clean metal substrate, and the isolated Bz molecule,
respectively. }
%Footnotes show the decomposition of the total PBE+vdW$^ {\rm surf}$ adsorption energy into the contributions from vdW energy
%(\textit{E}$_{\rm ad}^{\rm vdW}$) and from PBE adsorption energy (\textit{E}$_{\rm ad}^{\rm PBE}$), based on the same
%PBE+vdW$^ {\rm surf}$ relaxed structure for each system. }
\begin{tabular}{clccc}
\hline
\hline
System&\multicolumn{1}{c}{Method}&\textit{E}$\rm_{ad}$  [eV] &\textit{d}$\rm_{CM}$  [{\AA}] & \textit{d}$\rm_{HM}$ [{\AA}]\\
\hline
\multirow{6}{*}{Bz/Pt(111)}&PBE+vdW$^{\rm surf}$&1.96      &2.08    &2.51\\
                           &optB88-vdW          &1.84      &2.12    &2.53\\
                           &vdW-DF              &0.77      &2.16    & 2.57  \\
                           &vdW-DF2             &0.34      &2.20    & 2.65  \\
                           &PBE&0.81&2.10&2.54\\
                           &LDA                 &2.30      &2.05    & 2.47 \\
                           &Experiment          &1.57-1.91\footnotemark[1] &2.02$\pm$0.02\footnotemark[2]& - \\
\hline
\multirow{6}{*}{Bz/Au(111)}&PBE+vdW$^{\rm surf}$&0.74     &3.05     &3.04\\
                           &optB88-vdW          &0.79     &3.23     &3.23\\
                           &vdW-DF              &0.59     &3.44     &3.42   \\
                           &vdW-DF2             &0.56     &3.29     &3.27   \\
                           &PBE&0.15&3.62&3.62\\
                           &LDA                 &0.49     &2.83     &2.82   \\
                           &Experiment          &0.73-0.87\footnotemark[3] &2.95-3.10\footnotemark[4]     & - \\
\hline
\hline
\end{tabular}
\footnotetext[1]{Heat of adsorption measured with calorimetry, at the same coverage (0.7 ML) used for the DFT calculations.~\cite{ihm2004}
The error estimates of $\pm$10\% are taken from the reference.~\cite{ihm2004} Recent work suggests reduced errors
of $\pm$5\%.~\cite{Campbell-2011}}
\footnotetext[2]{LEED experiment.~\cite{wander1991}}
\footnotetext[3]{TPD experiment.~\cite{victor2012,syomin2001}}
\footnotetext[4]{Deduced data based on the experimental workfunction for Bz on Au(111) and adsorption
distance for pentacene on Au(111).~\cite{toyoda2009,abad2009,abad2011}}
%\footnotetext[1]{\textit{E}$_{\rm ad}^{\rm vdW}$ = 1.17 eV, and \textit{E}$_{\rm ad}^{\rm PBE}$ = 0.79 eV for Bz/Pt(111).}
%\footnotetext[2]{\textit{E}$_{\rm ad}^{\rm vdW}$ = 0.80 eV, and \textit{E}$_{\rm ad}^{\rm PBE}$ = -0.06 eV for Bz/Au(111).}
\label{table1}
\end{table}
%%%%%%%%%%%%%%%%%%%%%%%%%%%%%%%%%%%%%%%%%%%%%%%%%%%%%%%%%%%%%%%%%%%%%%%%%%%%%%%%%%%%%%%%%%%%%%%%%%%%%%%%%%%%%%%%%%%%%%%%%%%%%%%%%%%%%%%%%%%%%%

%%%%%%%%%%%%%%%%%%%%%%%%%%%%%%%%%%%%%%%%%%%%%%%%%%%%%%%%%%%%%%%%%%%%%%%%%%%%%%%%%%%%%%%%%%%%%%%%%%%%%%%%%%%%%%%%%%%%%%%%%%
\begin{figure*}
\includegraphics[width= 15 cm]{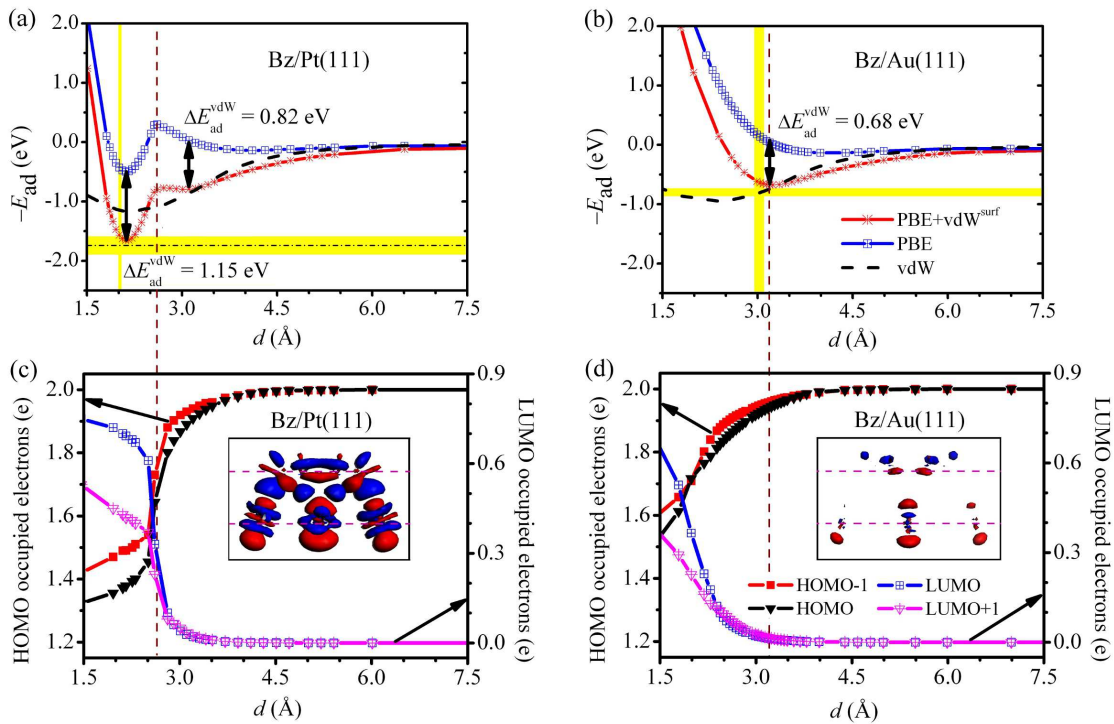}
\caption{\label{fig:wide} Top: Adsorption energy --\textit{E}$\rm _{ad}$ as a function of the adsorption height \textit{d} for Bz on Pt(111)
(a) and on Au(111) (b) from the PBE and PBE+vdW$\rm ^{surf}$ methods (the carbon backbone height $d$ from the surface is kept fixed).
%Similar curves are obtained from the optB88-vdW method, see supplemental material.
The experimental binding distances and adsorption energies are indicated by yellow intervals. Bottom: Integrated projected molecular density of states~\cite{modos2008} for the HOMO-1, HOMO, LUMO, and LUMO+1 orbitals of the benzene molecule as a function of $d$ for Bz on Pt(111) (c) and on Au(111) (d). The inset in panel (c) shows a side view of the electron density difference, which was obtained by subtracting electron density of isolated molecule and clean surface from an electron density plot of the entire adsorbed system, upon
Bz adsorption on Pt(111) at $d$=2.08 {\AA} (red = electron depletion, blue = electron accumulation).
For the same value of the isosurface (0.04 {\AA}$^ {-3}$), the electron density difference for Bz/Au(111)
at $d$=2.08 {\AA} is significantly weaker, see the inset in panel (d).}
%The binding curve is calculated by changing the adsorption height $d$ of Bz,
%which is evaluated relative to the position of the unrelaxed topmost metal layer. For each structure, we fix the \textit{z} coordinates of
%the carbon backbone and the metal atoms in the bottommost four of the employed six layer surface model.}
\end{figure*}
%%%%%%%%%%%%%%%%%%%%%%%%%%%%%%%%%%%%%%%%%%%%%%%%%%%%%%%%%%%%%%%%%%%%%%%%%%%%%%%%%%%%%%%%%%%%%%%%%%%%%%%%%%%%%%%%%%%%%%%%%%

The analysis of the equilibrium distances and adsorption energies in Table I demonstrates that both PBE+vdW$^{\rm surf}$ and optB88-vdW methods
lead to an excellent agreement with the available experimental data.
For the Bz/Pt(111) system, the PBE+vdW$^{\rm surf}$ adsorption energy of 1.96 eV is close to that from optB88-vdW (1.84 eV)
and both methods agree with the measured calorimetry values at 0.7 ML (1.57-1.91 eV, the same coverage used for DFT calculations).~\cite{ihm2004}
The PBE+vdW$^{\rm surf}$ adsorption energy converges to 2.18 eV with increasing surface cell size,
within the error bar of calorimetry measurements in the limit of zero coverage (1.84-2.25 eV).~\cite{ihm2004}
Note that the exclusion of the vdW interactions in the \textit{strongly adsorbed} Bz/Pt(111) system would lead to a significant
reduction in the binding energy (0.81 eV from PBE), in disagreement with the experimental data.
The adsorption energies computed using the vdW-DF and vdW-DF2 methods are even smaller than those calculated with PBE.
The adsorption energy for Bz adsorbed on the Au(111) surface is considerably smaller than that of Bz/Pt(111).
Also for Bz/Au(111), the PBE+vdW$^{\rm surf}$ adsorption energy (0.74 eV) agrees very well with both the optB88-vdW result (0.79 eV)
and the experimental temperature-programmed desorption (TPD) data at 0.1 ML (0.73-0.87 eV).~\cite{victor2012,syomin2001}
We conclude that PBE+vdW$^{\rm{surf}}$ and optB88-vdW methods yield quantitative
agreement with experimental adsorption distances and energies for both weakly and
strongly adsorbed Bz molecule. In contrast, LDA calculations are not systematic,
underbinding for Au(111) and overbinding for Pt(111).
%In contrast, vdW-DF and vdW-DF2 methods
%fail to provide even a qualitatively correct description of Bz/Pt(111),
%while LDA is not systematic, yielding underbinding for Au(111) and
%overbinding for Pt(111).

Deeper insight into the mechanism of Bz adsorption can be gained upon analysis of the binding energy curves, \textit{E}$\rm{_{ad}}$(\textit{d}), in Fig. 2.
The binding energy curves exhibit several characteristic effects. With decreasing distance the binding energy of the adsorbate system increases,
determined mainly by vdW interactions,
and here (for \textit{d}$>$3.5 {\AA}) Au(111) and Pt(111) show very similar behavior. In both cases the calculations show a small broadening of the energy levels.
The fully occupied \textit{d}-band of Au is obviously stiffer than the partially empty \textit{d}-band of Pt. In fact, for the latter the Pauli repulsion can be weakened
by the rearrangement of \textit{d}-electron density (a similar effect has been investigated in previous literature~\cite{silva2003}). As a consequence, the Bz molecule gets closer
to the surface of Pt and the HOMO and LUMO levels of the combined system broaden and hybridize noticeably. This goes together with significant electron transfer:
The HOMO and HOMO-1 orbitals of Bz molecule get partially depleted and the LUMO and LUMO+1 orbitals become partially filled. This behavior (broadening, shift, hybridization of levels, and electron transfer)
is a clear signature of the covalent interaction for Bz/Pt(111). Thus, at the adsorption geometry the wave-function has attained a qualitatively new character.
Figure 2 shows that this character change sets in for Pt at a distance of 3.1 {\AA}. At 2.6 {\AA} nearly a full electron has been transferred from the HOMO and HOMO-1 levels to
the LUMO and LUMO+1 levels, and in the total energy we observe a ``phase transition behavior'' (cf. the peak at 2.6 {\AA}). Finally, at the equilibrium geometry the
electron transfer (rearrangement) is as large as $\sim$1.1 electrons.
For Au surface the process is much weaker and -- not surprisingly -- a covalent contribution to the adsorption
process remains negligible. Thus, the vdW attraction governs the interaction.

Further inspection of the electron density difference at the strongly bound minimum for
Bz/Pt(111) in Fig. 2(c) demonstrates the rather strong hybridization between the HOMO/LUMO
orbitals of Bz and the $d_{z^2}$ orbitals of the Pt(111) atoms.
For the same adsorption height, the electron density difference for Bz/Au(111) is weak (see Fig. 2(d), inset).
The presence of two minima for Bz/Pt(111) resembles the recently
studied bonding of graphene on Ni(111).~\cite{mittendorfer2011,li2012}
However, the adsorption of Bz on Pt(111) exhibits a different feature.
In fact, Bz is exothermically bound on Pt(111) already when using PBE
without vdW interactions, while the PBE adsorption energy is endothermic for graphene on Ni(111).
Evidently, the functionalization of aromatic molecules would allow to control the
position and stability of the two adsorption minima on metallic surfaces.

Interestingly, while covalency is crucial for the Bz/Pt(111) bonding character, energetically the vdW contribution is in fact significant.
Upon inclusion of vdW interactions, the binding behavior is strongly modified -- the barrier to adsorption vanishes, and a precursor
physisorption state emerges for Bz/Pt(111).
The PBE+vdW$^{\rm{surf}}$ method lowers the adsorption energy from 0.50 eV (pure PBE value) to 1.65 eV in Fig. 2. Thus the final adsorption
results from a strongly concerted, synergistic effort.  Upon comparing the binding curves for Bz/Pt(111) and Bz/Au(111) we see that
the vdW contribution (due to vdW$^{\rm{surf}}$) for Bz/Pt(111), 1.15 eV,  is even stronger that that for Bz/Au(111), 0.68 eV.
The screened Bz/surface $C_3$ vdW coefficient is essentially the same for Pt(111) and Au(111) surfaces (2.17 and 2.02 hartree bohr$^3$, respectively).
Therefore, we conclude that the larger contribution of the vdW energy in the case of covalent bonding comes from the rather short adsorption
distance of the Bz molecule from the surface.

Our conclusions hold in general for the adsorption of Bz on other transition metal surfaces.
For Bz/Ir(111), the binding curve shows the same characteristic features as for Bz/Pt(111) in Fig. 2.
For Bz adsorbed on the Pd(111), Rh(111), and Ir(111) surfaces, the vdW energy contributions from the PBE+vdW$^ {\rm surf}$ method are in the range of 0.97-1.21 eV,
greater than those for Bz physisorbed on Ag(111) and Au(111) (0.68-0.82 eV).
Even larger vdW energies are found in more complex polyaromatic adsorption systems. For instance, the vdW energy is determined to be 1.77 eV
for naphthalene (C$_{10}$H$_8$) on the Pt(111) surface with (5 $\times$ 4) unit cell. Also for this case the calculated adsorption energy
from PBE+vdW$^{\rm surf}$ (2.91 eV) is within the experimental error bars (2.80-3.42 eV).~\cite{gottfried2006}
For anthracene (C$_{14}$H$_{10}$) on the Pt(111) surface with (6 $\times$ 4) unit cell,
the adsorption energy contributed by vdW interactions (2.42 eV) largely exceeds that
determined from the PBE functional (1.38 eV).

%%%%%%%%%%%%%%%%%%%%%%%%%%%%%%%%%%%%%%%%%%%%%%%%%%%%%%%%%%%%%%%%%%%%%%%%%%%%%%%%%%%%%%%%%%%%%%%%%%%%%%%%%%%%%%%%%%%%%%%%%%%%%
%\Other systems with covalent bonds.
\begin{table}
\centering
\caption{Adsorption energies \textit{E}$\rm_{ad}$ (eV) of Bz adsorbed on (111) surfaces of Ag, Pd, Rh, and Ir.}
%Footnotes show the decomposition of the total PBE+vdW$^ {\rm surf}$ adsorption energy into the
%contributions from vdW energy (\textit{E}$_{\rm ad}^{\rm vdW}$) and from PBE adsorption energy
%(\textit{E}$_{\rm ad}^{\rm PBE}$), based on the same PBE+vdW$^ {\rm surf}$ relaxed structure for each system.}
\begin{tabular}{ccccc}
\hline
\hline
System&PBE&PBE+vdW$^{\rm surf}$&optB88-vdW\\
\hline
Bz/Ag(111)& 0.09 & 0.75 & 0.72\\
Bz/Pd(111)&1.17&2.14&1.91\\
Bz/Rh(111)&1.48&2.52&2.27\\
Bz/Ir(111)&1.10&2.24&2.09\\
%Naphthalene/Pt(111)&1.17&- &2.91\\
%Anthracene/Pt(111)&1.37&- &3.80\\
\hline
\hline
\end{tabular}
%\footnotetext[1]{\textit{E}$_{\rm ad}^{\rm vdW}$ = 0.97 eV, and \textit{E}$_{\rm ad}^{\rm PBE}$ = 1.17 for Bz/Pd(111).}
%\footnotetext[2]{\textit{E}$_{\rm ad}^{\rm vdW}$ = 1.21 eV, and \textit{E}$_{\rm ad}^{\rm PBE}$ = 1.31 for Bz/Rh(111).}
%\footnotetext[3]{\textit{E}$_{\rm ad}^{\rm vdW}$ = 1.14 eV, and \textit{E}$_{\rm ad}^{\rm PBE}$ = 1.10 for Bz/Ir(111).}
\label{table2}
\end{table}
%%%%%%%%%%%%%%%%%%%%%%%%%%%%%%%%%%%%%%%%%%%%%%%%%%%%%%%%%%%%%%%%%%%%%%%%%%%%%%%%%%%%%%%%%%%%%%%%%%%%%%%%%%%%%%%%%%%%%%%%%%%%%%%%%

\section{CONCLUSION}

In summary, we have demonstrated that the inclusion of vdW interactions qualitatively changes
the adsorption behavior for benzene strongly interacting with (111) metal surfaces.
The vdW energy in Bz/Pt(111), a typical strongly adsorbed system, is almost 0.5 eV greater
than that in Bz/Au(111), a typical physisorbed system.
The bonding mechanism of Bz/Pt(111) stems from a synergistic effort of covalent bonding
and vdW interactions, and it is characterized by a peculiar ``phase transition''
behavior in the projected HOMO/LUMO occupations of the benzene molecule.
Our findings for Bz adsorbed on Ag, Au, Pt, Pd, Rh, and Ir surfaces
indicate that DFT calculations with dispersion interactions are essential for both weakly and strongly bound
molecules on surfaces.

%%%%%%%%%%%%%%%%%%%%%%%%%%%%%%%%%%%%%%%%%%%%%%%%%%%%%%%%%%%%%%%%%%%%%%%%%%%%%%%%%%%%%%%%%%%%%%%%%%%%%%%%%%%%%%%%%%%%%%%%%%
\section*{ACKNOWLEDGMENTS}

We are grateful for support from the FP7 Marie Curie Actions of the European Commission, via the Initial Training Network
SMALL (MCITN-238804). W.L. was funded by a fellowship from the Alexander von Humboldt Foundation.
A.T. acknowledges support from the European Research Council (ERC Starting Grant \texttt{VDW-CMAT}).
A.M. is supported by the ERC (ERC Starting Grant \texttt{QUANTUMCRASS}) and by the Royal Society through a
Wolfson Research Merit Award. J.C. is a Ram{\'o}n y Cajal fellow supported by the Spanish Government.

%\bibliography{Benzene}
%\bibliographystyle{apsrev4-1}
%Merlin.mbs v4.21 2009-07-09.
%

%%%%%%%%%%%%%%%%%%%%%%%%%%%%%%%%%%%%%%%%%%%%%%%%%%%%%%%%%%%%%%%%%%%%%%%%%%%%%%%%%%%%%%%%%%%%%%%%%%%%%%%%%%%%%%%%%%%%%%%%%

\end{document}